\documentclass[amsmath,amssymb,prl]{revtex4}

\usepackage{graphicx}
\usepackage{dcolumn}
\usepackage{bm}

\newcommand{\ket}[1]{\left|#1\right\rangle}
\begin{document}

\title{Universal quantum computation with the Orbital Angular Momentum of a single photon}
\author{Juan Carlos Garc\'ia-Escart\'in}
\email{juagar@tel.uva.es}
\author{Pedro Chamorro-Posada}
\affiliation{Dpto. de Teor\'ia de la Se\~{n}al y Comunicaciones. ETSI de Telecomunicaci\'on. Universidad de Valladolid. Campus Miguel Delibes. Paseo Bel\'en 15. 47011 Valladolid. Spain.}
\date{\today}
\begin{abstract}
We prove that a single photon with quantum data encoded in its orbital angular momentum can be manipulated with simple optical elements to provide any desired quantum computation. We will show how to build any quantum unitary operator using beamsplitters, phase shifters, holograms and an extraction gate based on quantum interrogation. The advantages and challenges of these approach are then discussed, in particular the problem of the readout of the results.
\end{abstract}

\maketitle

\section{Optical implementations of quantum computing}
\label{optical}
Quantum information processing offers a new model of computation and communications. Certain tasks which can only be performed with a limited efficiency in a classical computer can be carried out efficiently in the quantum case \cite{NC00}. For that reason, there is a great interest in finding a suitable physical implementation of a quantum computer. 

The DiVincenzo criteria give an important guide to the conditions a physical realization of a quantum computer should meet \cite{Div00}. A practical quantum computer should be implemented over a system which is scalable with the input size and can be easily initialized and read. Additionally, the system must be able to carry out any desired quantum operation. This means that we must be able to implement any logic function and that the coherence time of the system (the time in which the quantum properties of the system are maintained) is long enough to finish the computation. 

Optical implementations seem particularly attractive. Photons can be initialized and read (generated and detected) with relative ease and have possibly the longest coherence time from all the quantum information candidates. They are also very well suited for communications. However, the interaction between different photons is complicated and needs to be mediated by non-linear processes. One proposed solution has been the use of measurement, like in the Linear Optics Quantum Computer of Knill, Laflamme and Milburn \cite{KLM01}.

Most notably, if we only have a single photon, there is always a way to perform any desired quantum computation with linear optics. A linear optics multiport can be described by a scattering matrix that gives the relationship between the amplitudes of the fields at the different input and output modes. For a single photon in $2^n$ input spatial modes, the scattering matrix corresponds to the unitary operator that gives the quantum evolution of a system of $n$ qubits ($n$ quantum information units). Any desired unitary operator of this kind can be implement using only beamplitters and phase shifters \cite{RZBB94}. Consequently, any desired quantum computation can be implemented on a single photon at the cost of having a number of paths which grows exponentially with the number of qubits of the input \cite{CAK98}. In this case, we can obtain universal logic but cannot meet the scalability requisite. 

In this Letter, we propose a compact variation of single photon quantum computation with orbital angular momentum encoding. We will show how using only a limited set of optical elements it is possible to provide any unitary operator without needing an exponential number of space paths. However, there will be a conflict with the readout criterion. We will discuss some possible ways out of this dilemma.

\section{OAM encoding}
\label{encoding}
An $n$-qubit quantum computer needs $2^n$ orthogonal states to encode all the possible values of $n$ bits as well as superposition states. Instead of the usual option of having $n$ two-level quantum systems, we will base our encoding on the orbital angular momentum (OAM) states of light. For a single photon, we can define orthogonal OAM states $\ket{\ell}$ that correspond to the photon carrying an OAM of $\ell\hbar$, for an integer $\ell$ \cite{ABP03}. A generic state will be of the form $\ket{\psi}=\sum_{\ell=0}^{2^n-1}\alpha_{\ell}\ket{\ell}$. The integer $\ell$ encodes the binary sequence $\ell=b_{n-1}b_{n-2}\cdots b_0$. These states can be produced efficiently \cite{MTT01}. The primary advantage of this encoding is its compactness. The same photon in the same spatial mode can be in a superposition of $2^n$ OAM modes. 

\section{Building blocks}
\label{blocks}
In our single photon computer we will use standard optical elements. The basic devices and operations are:
\begin{itemize}
\item \emph{Phase shifters}: A phase shifter will introduce the same phase shift on all input states crossing it so that $PS_{\Phi}\ket{\ell}=e^{i\Phi}\ket{\ell}$.
\item \emph{Holograms}: Different holograms can be designed to produce an increase or decrease of the OAM index $\ell$ \cite{ADA98}. We will represent this operation as $H_k\ket{\ell}=\ket{\ell+k}$, where $k$ is an integer. 

\begin{figure}[ht!]
\includegraphics{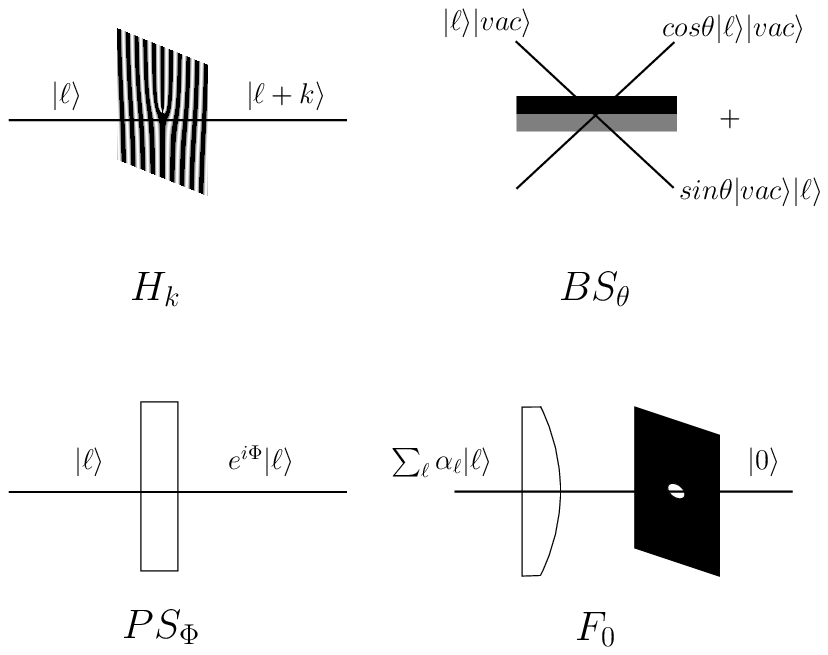}
\caption{\label{els} Representation of the optical elements.}
\end{figure}

\item \emph{Beamsplitters}: The different OAM inputs of different $\ell$ will be orthogonal and will not interfere in a beamspliter. We will have the same operation as in a single photon for each of the OAM states. As we only have one photon,  all the inputs will have an OAM state and the vacuum state $\ket{vac}$. The evolution is given by $BS_{\theta}\ket{\ell}\ket{vac}=\cos \theta \ket{\ell}\ket{vac}+\sin \theta \ket{vac}\ket{\ell}$ and $BS_{\theta}\ket{vac}\ket{\ell}=\sin \theta \ket{\ell}\ket{vac}-\cos \theta \ket{vac}\ket{\ell}$. Figure \ref{els} shows our representation for the beamsplitter. We will consider that reflection on the grey side comes with a phase shift. This sign choice is only a matter of reference \cite{SGL04}. 

\item \emph{OAM filters}: We will need filters that can absorb all the photons but those of a particular OAM. Although it might be more efficient to describe tailor-made filters for any $\ket{\ell}$ states, for our proof of principle we only need filters for the $\ket{0}$ state. Using holograms we can produce any other necessary filter. The filter will be a non-linear element. We have the evolution $F_0\sum_{\ell=0}^{2^n-1}\alpha_{\ell}\ket{\ell}=\ket{0}$ with a probability $|\alpha_0|^2$ and the photon is absorbed with probability $1-|\alpha_0|^2$. There can be different ways to implement this kind of filter. We will employ a pinhole filter. Only the $\ket{0}$ OAM state has a planar phase front which can be focused through a pinhole. All the other modes will be blocked by the screen. Another option is coupling the photon into a single mode optical fibre (SMF). This coupling will only be efficient for the $\ell=0$ mode \cite{MVW01}.
\end{itemize}

Figure \ref{els} shows the representation of all these elements. 

\subsection{The extraction gate}
\label{extraction}
The main element of our construction is a gate that is able to separate a state of a given value of OAM from a superposition and take it to a different spatial mode. The operation can be seen as: $E_m(\sum_{\ell=0}^{2^n-1}\alpha_{\ell}\ket{\ell}\ket{vac})=\sum_{\ell\neq m}\alpha_{\ell}\ket{\ell}\ket{vac}+\alpha_m\ket{vac}\ket{0}$. We have made the extracted state to be $\ket{0}$ because of the implementation method and its use in the proof, but a suitable hologram can produce the more natural gate in which the $\ket{m}$ state is extracted.

\begin{figure}[ht!]
\includegraphics{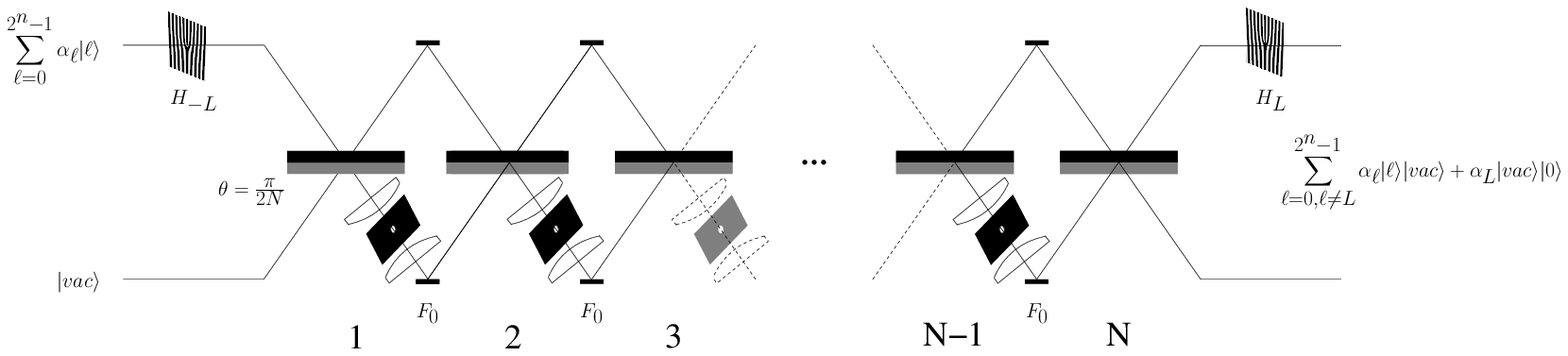}
\caption{\label{extractor} Extractor gate setup. After the desired state has been selected, a combination of filtering and measurement produces a separation in space.}
\end{figure}

Our extraction gate is based on Quantum Interrogation, where frequent measurement results in a directed quantum evolution \cite{KWH95}. Figure \ref{extractor} shows the basic setup. At the input, we have a $-m$ hologram. From that point, we can study the evolution for two cases, the $\ket{0}$ state (the former $\ket{m}$ state) and all the other $\ket\ell$ states. The filter separates the evolution of these two group of states. 

Take a generic $\ket{\ell}$ state with $\ell\neq 0$ for the input. The first beamsplitter will have at its output the state $\cos\theta \ket{\ell}\ket{vac}+\sin\theta\ket{vac}\ket{\ell}$. The second spatial mode is then directed to the filter. At the filter, the state is projected to $\ket{\ell}\ket{vac}$ with probability $\cos^2\theta$. This will happen at each of the $N$ beamsplitters. For our angle $\theta=\frac{\pi}{2N}$ the final state will be $\ket{\ell}\ket{vac}$ with a probability $\cos^{2N}\theta\approx1-\frac{\pi^2}{4N}$ which can be made arbitrarily close to 1. If the photon were absorbed, the gate fails and we would have to start the computation again.

The filter has no effect on the $\ket{0}$ OAM state. The input sees $N$ beamsplitters so that $E_m\ket{m}\ket{vac}=BS_{\theta}^N\ket{0}\ket{vac}=\ket{vac}\ket{0}$. To prove this we only need to notice that the beamsplitter matrix is a rotation by an angle $\theta$ and that $BS_{\theta}^N=BS_{N\theta}=BS_{\frac{\pi}{2}}$.

The conjuction of both evolutions gives us the desired operation. Notice that, despite the non-linear measurements, the gate is reversible. If we repeat the same configuration at the output of an extraction gate, the extracted term will be reintegrated into the superposition. 

There are two practical details worth commenting. First, we have introduced lenses that undo the focusing through the pinhole so that a proper interference occurs at all the necessary points. Second, as reflections take the OAM state from $\ket{\ell}$ to $\ket{-\ell}$, we need to make sure that all the paths have an even total number of reflections. 

\section{Universality}
\label{universal}
To prove that these gates can be used to implement any unitary operation we only need to show we can build two-level unitary matrices $U_{m,n}$ \cite{RZBB94}. The $U_{m,n}$ operation only affects states of OAM $m\hbar$ and $n\hbar$ so that $U_{m,n}\ket{m}=u_{m,m}\ket{m}+u_{m,n}\ket{n}$, $U_{m,n}\ket{n}=u_{n,m}\ket{m}+u_{n,n}\ket{n}$ and $U_{m,n}\ket{\ell}=\ket{\ell}$ for $\ell\neq m,n$. The matrix
\begin{center} 
$U_{2}^{m,n}= \left( \begin{array}{rr} u_{m,m} & u_{m,n} \\ u_{n,m} & u_{n,n} \end{array} \right).$ \end{center}
must be unitary. 

\begin{figure}[ht!]
\includegraphics{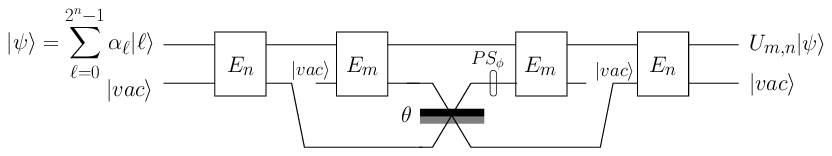}
\caption{\label{twolevel} Implementation of a two level unitary for OAM encoding.}
\end{figure}

Figure \ref{twolevel} shows a simple OAM implementation based on the extraction gate. First, we extract the $\ket{m}$ and $\ket{n}$ terms of the input state and take them to the $\ket{0}$ OAM mode of two separate spatial modes. Then, using linear optics elements (a beamsplitter and a phaseshifter), we can give any $2\times2$ unitary \cite{RZBB94}. Finally, the new superposition is integrated back into the spatial mode of the rest of the OAM states. The resulting evolution is
\begin{equation}
\sum_{\ell=0}^{2^n-1}\alpha_{\ell}\ket{\ell}\rightarrow\sum_{\ell\neq m,n}\alpha_{\ell}\ket{\ell}+(u_{m,m}\alpha_m+u_{m,n}\alpha_n)\ket{m}+(u_{n,m}\alpha_m+u_{n,n}\alpha_n)\ket{n},
\end{equation}
which corresponds to the desired two-level unitary.

This completes the proof that OAM encoding is enough to have a space-efficient universal set of gates.


\section{The readout problem}
\label{readout}
We have seen we can perform any desired quantum computation with the OAM of a single photon. However, the computation cannot be said to have finished until we have read the solution from the resulting OAM state. OAM sorters are devices that can discriminate between different OAM states of a single photon \cite{LPB02}. In the sorter, the photon is directed through different paths depending on its OAM. There are $n$ binary decision points in a branching optical setup with $2^n$ arms. This branching takes us back to an exponential growth problem. Alternative approaches to OAM sorting share this exponential growth problem. In them, we either have an increasing number of spatial paths or we need to discriminate between an exponentially high number of frequencies \cite{XWK01,VTP03}. This limitation is a consequence of the destructive nature of photon measurement. A series of $n$ quantum nondemolition measurements of the photon would determine the OAM state efficiently, but these measurements are as experimentally challenging as achieving interaction between photons and, in many respects, are equivalent to them. 

There are different solutions to this problem. On one hand, if we have an algorithm where the result is always the same, or the same with very high probability, like in a simple Grover search, we can run the single photon computer $n$ times. In that case $n$ binary decisions are enough to determine all the bits of the output. At each measurement we discover the branch the photon would have taken in the whole sorter setup. With the $n$ measurements, we can learn each of the $n$ bits encoded in the OAM state using only one path separation operation. 

If this is not the case, we can use OAM demultiplexers. OAM demultiplexers can take the information from the OAM of a single photon and put it into the path of $n$ photons \cite{GC08}. In this new encoding, readout is efficient. In order to build an OAM demultiplexer we would need $2n-1$ path CNOT gates, which are the kind of gate that requires the strong photon-photon interaction we are trying to avoid. Nevertheless, this would only impose a \emph{fixed price for any arbitary computation} which only grows linearly with the size of the input. This can be an important saving when compared with alternative optical models. 

\section{Discussion}
\label{discussion}
We have presented a scheme for universal computation using the OAM of a single photon. The quantum gates are compact in space, but the readout of the final state requires either repeated runs of the same computation or the use of a limited number of optical CNOT gates.

There are a few challenges to an efficient experimental realization. The main obstacle is the number of elementary gates needed in a concrete implementation. Although we have given a proof of universality, the number of optical gates used to implement a particular operation can be exponentially high. All universal sets of gates have this problem, but, in many implementation proposals, there are interesting circuits (for instance circuits for Shor's or Grover algorithms) which are efficient in the number of elementary gates. For OAM it would be an important advance to find a simple optical implementation for a ``dispersing'' gate, like the Hadamard gate or the Quantum Fourier Transform, which appear in many quantum algorithms which offer a speedup with respect to classical ones. In these gates, the new probability of each state comes from the interference of the probability amplitudes of most of the possible logical states. An element that takes a photon from any single OAM state into a superposition of many others would be an important step towards useful single photon quantum computers with OAM. 

The proof that universal logic is, indeed, attainable with an OAM encoding, suggests it is worth a further investigation into this new kind of optical quantum computer. The experimental realization of a single photon quantum computer, at least for certain tasks, seems feasible, although problems like overcoming losses in the optical elements, achieving interfereometric stability in the extraction gate and generating and processing states of a high OAM can limit the practical implementation.

\section{Acknowledgements}
This research has been funded by Junta de Castilla y Le\'on grant No. VA001A08 and MICINN and FEDER project TEC2007-67429-C02-01.

\newcommand{\noopsort}[1]{} \newcommand{\printfirst}[2]{#1}
  \newcommand{\singleletter}[1]{#1} \newcommand{\switchargs}[2]{#2#1}

\end{document}